\documentclass{article}

\usepackage{booktabs} % For formal tables
\usepackage{xspace}
\usepackage{subfigure}
\usepackage[utf8]{inputenc}
\usepackage{graphicx}
\usepackage{authblk}

\begin{document}

\title{{\sc Traj2User}: exploiting embeddings for computing similarity of users mobile behavior}

 \date{}

\author[1]{Andrea Esuli}
\author[2]{Lucas May Petry}
\author[1]{Chiara Renso}
\author[2]{Vania Bogorny}                            
\affil[1]{ISTI-CNR, Pisa, Italy}
\affil[2]{Programa de Pós-Graduação em Ciência da Computação, Universidade Federal de Santa Catarina, Florianopolis, Brazil}

\maketitle

\begin{abstract}
Semantic trajectories are high level representations of user movements where several aspects related to the movement context are represented as heterogeneous textual labels.
With the objective of finding a meaningful similarity measure for semantically enriched trajectories, we propose  \textsc{Traj2User}, a \textsc{Word2Vec}-inspired method for the generation of a vector representation of user movements as {\em user embeddings}.

\textsc{Traj2User} uses simple representations of trajectories and delegates the definition of the similarity model to the learning process of the network.
Preliminary results show that \textsc{Traj2User} is able to generate effective user embeddings.
\end{abstract}

\section{Introduction}
\label{sec:introduction}
The widespread use of GPS-equipped smartphones or positioning sensors applied to vehicles and animals, tend to produce a high number of trajectories, recording the spatio-temporal evolution of these objects.
These raw trajectories can be enriched with semantic information to what is called {\em semantic trajectories} \cite{spaccapietra2008,SurveyACM}, adding more meaning to the pure geometric movement facets.

In the era of Big Data, with the explosion of geolocated social media and other kinds of user generated data (e.g. Wikipedia, Flickr, etc), human mobility data  can be significantly enriched with  information that encompasses our daily life.
Enriching information include  weather conditions, the  transportation means, the goal or the activity performed during the movement, the opinions and comments about people and places, the mood, being with a friend, just to name a few examples discussed 
in \cite{Ferrero2016} and \cite{constant2013}. 

Being able to find similarities between trajectories enables several analysis methods like clustering or applications like recommendation systems. 
Several similarity measures have been proposed for 
both raw and semantic trajectories, such as  EDR\cite{Chen2005}, LCSS \cite{Vlachos2002}, Cats \cite{hung2015}, EDWp \cite{ranu2015},  UMS\cite{furtado2017}, MSM \cite{furtado16}, and other approaches as presented in \cite{berndt1994using}, \cite{Liu2012} and \cite{Xiao2014}. However, these methods analyze a few trajectory attributes and 
are far from considering all the different semantic aspects that involve movement: previous works have mainly analyzed semantic trajectories over one single aspect at a time, such as stops and moves, or transportation means, or  activities. 
 
The great challenge here is how to integrate all such heterogeneous dimensions in a similarity measure  dealing with space, time, and multiple semantics.

\begin{figure}
\centering
\includegraphics[width=0.6\columnwidth]{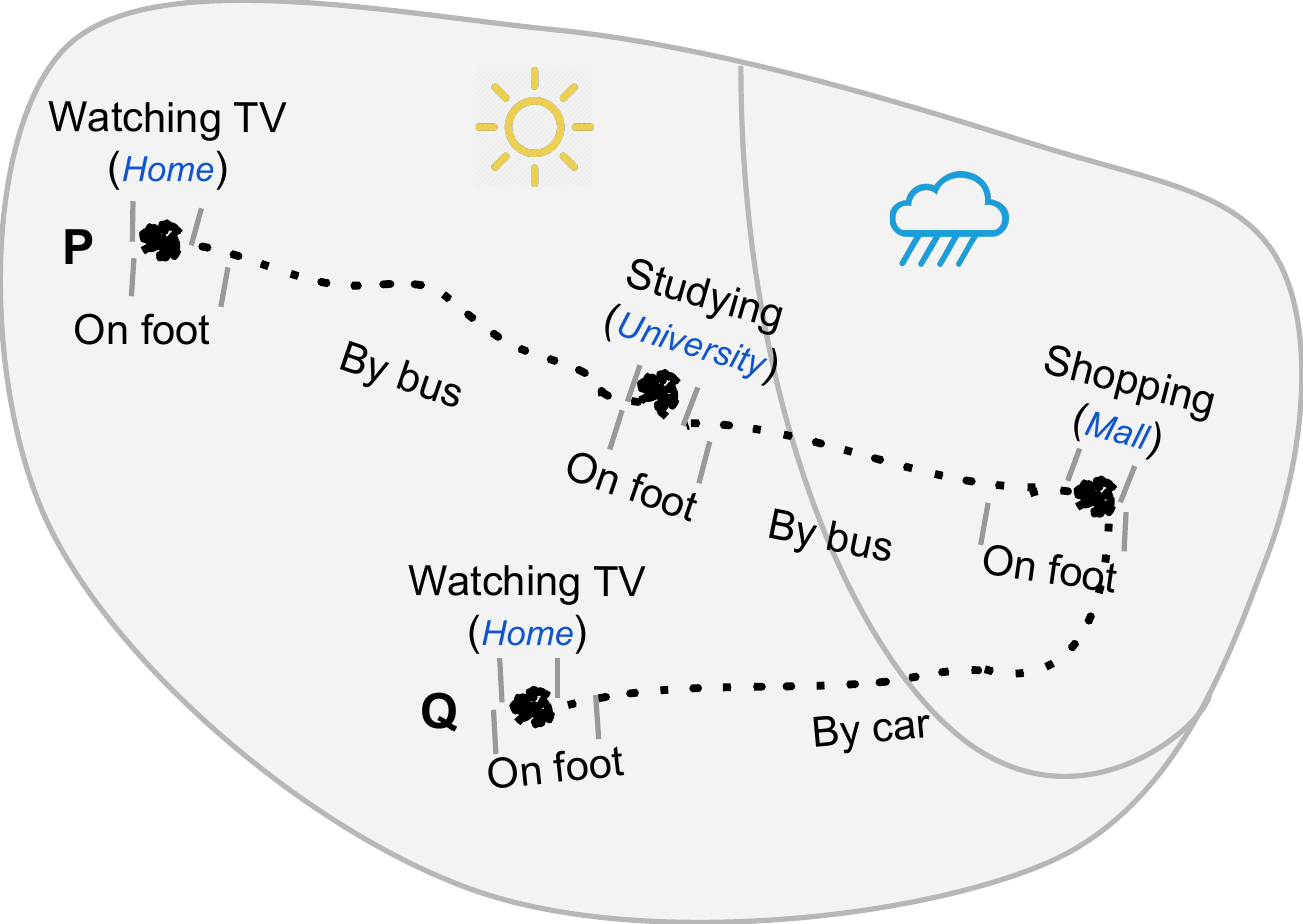}
\caption{\label{fig:trajectory} Example of two semantically rich trajectories P and Q.}
\vspace{-1.7em}
\end{figure}

Figure \ref{fig:trajectory} shows an example of two semantic trajectories $P$ and $Q$ from users $u_1$ and $u_2$. The question we want to answer is "{\em  how similar are $u_1$ and $u_2$ given their semantic trajectories $P$ and $Q$?}".  

As can be observed, first of all, the size of trajectories $P$ and $Q$ is different. Both users are \emph{shopping} at the same place (same spatial location) while \emph{on foot} (same transportation means) and when the weather is rainy (same weather condition). On the other hand, both users at some time (time dimension) watch TV, but on a different spatial location. Indeed, while the user of trajectory $P$ moves on foot and by bus, the user of trajectory $Q$ moves on foot and by car. 

Given all these different and heterogeneous data dimensions related to trajectories, where each dimension has its own similarity model, how can we compute the user similarity based on their trajectories $P$ and $Q$ considering all this information together? 

In this paper we represent users  by modeling the many semantic aspects that describe their movement habits. 
We take inspiration from {\em word embeddings} \cite{w2v2013} methods that model the semantics of a word, and its similarity with other words, by observing the many contexts of use of the word in the language.

We therefore introduce here, for the first time, the new concept of \emph{user embeddings} as a way to represent the semantically rich movements of a user. We also propose the \textsc{Traj2User} model for measuring the semantic trajectory similarity of moving users. 
%A peculiarity 
The main contribution of \textsc{Traj2User} is that it does neither need any explicit definition of the similarity functions for the data dimensions nor any explicit modeling of the relations between data dimensions, since these are implicitly learned from data.

Experiments show that the neural model of \textsc{Traj2User} learns user embeddings that better capture user similarity than other embedding methods that are also inspired to language models.
We believe that exploiting vectors to represent semantically complex movements is promising for the analysis of semantically enriched movement since several heterogeneous semantic aspects are uniformly modeled into a unique and compact form. One recent approach that considers Point of Interest (POI) and embeddings is the paper \cite{Poi2Vec}
where authors propose a
method to jointly model the user preference and the sequence of POIs for predicting future visitors for a given POI.

\section{Methodology}
\label{sec:methodology}

\subsection{Basic definitions}
We start our definitions with a state-of-art concept of raw trajectories, that has only space and time dimensions.

\noindent{\bf Definition}
(Raw trajectory) A raw trajectory is the sequence of timestamped locations of the traced moving object $o$ in the form $<o, <x_1,y_1,t_1, \ldots x_n,y_n,t_n>>$, where each $x_i,y_i$ represents the geographical coordinates and $t_i$ the timestamp for each  $i=1, \ldots, n$. 

In this paper 
we define a semantic trajectory based on trajectory segments (or movements).

\noindent{\bf Definition}
(Segmented Trajectory) A segmented trajectory of the moving object $o$ is a pair $t = <o, {s_1,...,s_n}>$ such that, for each $i=1,...,n$, $s_i$ is a contiguous part of the trajectory split based on some criteria.

Examples of segmenting criteria are the stops and moves \cite{spaccapietra2008},  
the transportation means  or the purpose of the trip \cite{constant2013}. 

\noindent{\bf Definition}
 (Semantic Trajectory) A semantic trajectory of an object $o$ is a pair $t\ =\ \ <o, \{<s_1,l_{11},..l_{1k}>,...,<s_n, l_{n1},... l_{nk}>\}>$ such that, for each i=1,\dots,n and j=1, \ldots, k $l_{ij}$ is the j-th semantic label for the segment $s_i$. 
 
A semantic trajectory, in the context of this work, is a segmented trajectory where each segment is enriched with a number of different labels that represent different semantic aspects of the trajectory segment. Considering again Figure \ref{fig:trajectory}, we see that the first trajectory has 5 segments while the second has 3 segments. The semantic representation of  the first trajectory is $P$ = $<$u$_1$, $<$s$_1$, "on foot", "watching TV","sunny"$>$, $<$s$_2$, "by bus, "going to University", "sunny"$>$, $<$s$_3$, "on foot", "studying", "sunny"$>$, $<$s$_4$, "by bus", "going shopping", "rainy"$>$, $<$s$_5$, "on foot", "shopping", "rainy"$>>$

\subsection{Building the user embeddings}

In this work we want to build vectorial representations in an embedding space, for each user, in a population of observed individuals.
Each user is represented by a variable-sized set of semantically rich movements.
We can consider each movement as a context of the user's life.
We observe that this situation is similar to the one in language modeling \cite{turney2010frequency,LeM14}, where one wants to model the semantic properties of a word and measure the semantic similarity between words by observing the many contexts in which the word appears.
The intuition of this work is based on the observation that the similarity between two words (e.g., king, queen) can be inferred by observing that they frequently appear in similar contexts (e.g., near other words such as castle, crown, empire, throne\ldots). Analogously, the similarity between two users can be inferred by observing that their semantic trajectories frequently have similar semantic values.

Following this parallel between words and users,  we make here a first exploration of how some methods for the construction of word embeddings can be applied to the process of constructing {\em user embeddings}.
In our approach we convert the label of each segment of a semantic trajectory into a vectorial representation (i.e., {\em a movement descriptor} as defined in Section \ref{sec:encoding}). We then combine the various movement descriptors of a user into a single {\em user embedding}.
We test three main models for the construction of user embeddings, which are described in Section \ref{sec:methods}.

\subsubsection{Encoding of trajectories into vectors}\label{sec:encoding}

The various labels of a semantic trajectory are heterogeneous by nature and therefore they may have very different forms.
In this work we decided to build a {\em movement descriptor} by encoding all the values of the segment labels using a {\em one-hot encoding}.
Given a label $a$ with $n^a$ possible different values, an actual value $l^a$ for the label is encoded as a $n^a$-sized vector with a {\em one} in the position corresponding to the value $l^a$ and $n^a-1$ {\em zeros} for all the other possible values.
The result of encoding a movement descriptor is thus a vector $d$ of length $|d| = \sum_{a\in A}n^a$, where $A$ is the set of labels. 

We are aware that one-hot encoding does not explicitly model any complex relation between values, e.g., values on ordinal scales.
Our approach follows the idea of leaving to the embedding generation method the burden of discovering the relations between the values of a label (and across labels).

\subsubsection{Generating user embeddings}\label{sec:methods}

The set of vectors representing movements and associated to a user must be reduced into a single vector that represents the user embedding.
A very simple approach is to sum all the vectors into a single vector, i.e., $e_i = \sum_{d_i\in{D_i}} d^i$, where $D^i$ identifies all the descriptors associated to the user $u_i$ (\textsc{Sum} method).

Stacking all the resulting sum vectors for all users produces a matrix $M$ of size $|U|\cdot |d|$.
This is similar to what is done when building a {\em word-context} matrix to create a language model, in which the position $T_{i,j}$ of a matrix $T$ of size $|V|\cdot|V|$ stores the sums up of how many times the word $w_j\in V$ appears nearby (in the context of) the word $w_i\in V$, where $V$ is a vocabulary.

Vectors deriving from raw sums can suffer skewness due to some values being much more frequent than others (e.g., weekday in our dataset).
Such very frequent values may dominate the directions of vectors and at the same time they may not be very discriminative.

Inspired by the approaches used in language modeling to address this issue, we tested a number of methods\footnote{We also tested simple $l^1$ and $l^2$ normalization, with negative results.} to produce a weight-corrected matrix $\hat{M}$ of user embeddings starting from the $Sum$ matrix\footnote{We also tested the application of such weighting methods to the distinct blocks of the matrix $M$ that identify the values from a single label, again with negative results.} $M$:

{\em Positive point-wise mutual information} (\textsc{PPMI}) \cite{Bullinaria2007} measures the dependence between the two observed variables (a user and a specific attribute value) only on the occurrence of events, i.e., how much the probability of the two events to occur together differs from chance\footnote{A small dataset, like our, can generate negative values due to lack of a sufficient number of observations. For this reason negative values are clipped to zero.}. 
\begin{equation}
\label{eq:ppmi}
\hat{M}_{(i,j)} = PPMI(u_i,d_j)=max(\log_2(\frac{P(u_i,d_j}{P(u_i)P(d_j)}),0)
\end{equation}

{\em Softmax} (\textsc{SM}) is often used to normalize a vector so that all its values are in the $[0,1]$ interval and their sum is one, i.e., a categorical probability distribution.
\begin{equation}
\label{eq:softmax}
\hat{M}_{(i,j)} = \sigma(M_{(i,j)}) = \frac{e^{M_{(i,j)}}}{\sum_{k=1}^{|d|} e^{M_{(i,k)}}}
\end{equation}

None of the methods described so far actually explores the latent correlations between the values of the labels, within a label or among labels.

Methods based on matrix decomposition, such as singular value decomposition (\textsc{SVD}) may exploit these latent relations modeling them in a project space that is not constrained to the original encoding of values.
Truncated SVD allows to generate shorter embeddings, possibly removing noise components from input data.
We tested SVD in combination with the previously described methods (i.e., \textsc{SVD-PPMI}, 

\textsc{SVD-SM})
with all the above listed methods,
and using various reduction factors $f$, where the resulting user embedding length is $|d|/f$.
In this paper we propose the \textsc{Traj2User} method that is inspired to the \textsc{Word2Vec} \cite{w2v2013} method for the generation of word embeddings.
The skip-gram variant of \textsc{Word2Vec} learns word embeddings as a by-product of training a two-layer network on the task of predicting from a single word other words that may appear in its context.

In the \textsc{Traj2User} network\footnote{The \textsc{Word2Vec} model uses many tricks, such as hierarchical softmax and negative sampling, to improve its performance.
In this work we explore a ``simple'' model, leaving the evaluation of these methods to future work.}, the role of the input word is taken by a user id, and the context by a movement descriptor. 
Given a user $u_i$ represented as a one-hot vector and one movement descriptors $d^i_j$, the first layer of the network selects the user embedding $e_i=W^Tu_i$, i.e., the matrix $W$ is the matrix of user embeddings.
The second layer multiplies $e_i$ to a second weight matrix $W'$ and applies the sigmoid activation function\footnote{We tested also the Softmax function, with negative results.} $S(x) = \frac{1}{1-e^{-x}}$ to predict the movement descriptor $\tilde{d}=S(W'^Te_i)$.
$\tilde{d}$ is compared to $d^i_j$ so that backpropagation updates $W'$ and $W$.

The length of user embeddings $|e_i|$ is a free parameter of the model.
Differently from the previous methods, \textsc{Traj2User} makes possible to set $|e_i| > |d|$, i.e., expanding the representation space.
The idea supporting the use of a representation space bigger than the original descriptors, is that it may be able to capture and represent more relationships between the attributes.

\section{Experiments}
\label{sec:experiments}

\subsection{Dataset}

\begin{table}

\begin{tabular}{l|p{0.75\linewidth}|r}
\bf Label  & \bf  Values & N.  \\ \hline \hline 
Purpose & at home, at work, eat out, shopping (food), shopping (other), social events, study, entertainment (sport, theater, museum\ldots), services (doctor, bank, hairdresser\ldots), picking up/taking someone home, fueling, stop (changing transportation), not specified, incomplete tracking (dead battery, app crash) & 14 \\\hline
Vehicle & car, bicycle, motorcycle, city public transport (but, metro), taxi, train, boat, on foot, not specified & 9 \\\hline
Start hour& 0-23 & 24 \\\hline
End hour & 0-23 & 24 \\\hline
Duration & $<$5 min, 5-8 min, 8-12 min, 12-20 min, $>$20 min & 5   \\\hline
Range & $<$1 km, 1 to 2 km, 2 to 4 km, 4 to 10 km,  $>10$ km & 5\\\hline
Weather & sunny, rain, fog, cloudy, not specified & 5  \\\hline
Weekday & weekday, weekend & 2 \\\hline
& \bf Total count & 88 \\
\end{tabular}
\caption{\label{tab:fields}Attributes describing trajectories in the dataset.}
\end{table}

The raw trajectories dataset has been collected by volunteers using GPS-enabled smartphones in the area of Pisa in the period from May 20, 2014 to September 30, 2014, in the context of the TagMyDay\footnote{We obtained this data under a non-disclosure agreement, therefore we cannot directly redistribute it. More information at http://kdd.isti.cnr.it/project/tagmyday} experiment.
Each user tracked his/her movements by using a GPS tracking {\em app} installed in their mobile phone. Each volunteer could freely decide which part of their daily movement to track by starting the application. Each trajectory has been uploaded into the experiment web site. 
Here, trajectory segments are automatically computed from the raw data identifying the different movements between two stops.
Then, from the web interface the users could annotate these segments with semantic labels like the purpose of the trip, the means of transportation, the weather. Other information has been computed automatically from the raw trajectory like the temporal duration of the segment, the spatial length and the day of the week (weekday, weekend). 

After the annotation task, a semantically enriched trajectory is identified by the attributes listed in Table \ref{tab:fields}.

The dataset contains traces of 157 users for a total of 10,880 segments.
The distribution of the number of trajectories associated to the users follows an exponential decay with the most active user having produced 727 segments and a tail of 39 users with less than 5 segments. 
The encoding of a trajectory segment into a movement descriptor produces a vector of length $|d|=88$ (see Table \ref{tab:fields}).

\subsection{Experiments and results}\label{sec:results}

\begin{table}
\centering
\begin{tabular}{|l|ccccc|}\hline
 & \multicolumn{5}{|l|}{compression factor $f$} \\\cline{2-6}
method & 0.5 & 1 & 2 & 4 & 8 \\\hline
\textsc{Sum} & n/a & 0.748 & n/a & n/a & n/a \\\hline
PPMI & n/a & 0.708 & n/a & n/a & n/a \\
SM & n/a & 0.378 & n/a & n/a & n/a \\\hline
SVD-PPMI & n/a & 0.708 & 0.706 & 0.666 & 0.605 \\
SVD-SM & n/a & 0.394 & 0.394 & 0.393 & 0.363 \\\hline
\textsc{Traj2User} & 0.858 & 0.858 & 0.844 & 0.803 & 0.774 \\\hline
\end{tabular}
\caption{\label{tab:results}Comparison of the methods for the generation of user embeddings. Average MRR value across 1,000 test pairs.}

\end{table}

We designed our experiment as a similarity search problem.
Given two users $u_a$ and $u_b$ which are known to have very similar mobility habits, we rank all users in $U\setminus \{u_a\}$ by the similarity of their embedding with $e_a$, using the cosine similarity function, and observe the rank $r^a_b$ of $u_b$.
We repeat this on a large set of pairs of similar users $P$ and measure the mean reciprocal rank $MRR(P) = \frac{1}{|P|}\sum_{(u_a,u_b)\in P}\frac{1}{r^a_b}$ across all pairs.
The higher the $MRR$ score the better, since it indicates that the embeddings capture the similarities among the $(u_a,u_b)$ pairs.

Our dataset does not have an explicit evaluation of similarities among users.
To solve this issue we created the pairs of similar users by randomly selecting a user $u_a$ from $U$ and then randomly distributing the movement descriptors of $u_a$ between two 'virtual' users $u_a'$ and $u_a''$, which actually define a test pair.
Given a pair $(u_a', u_a'')$, to capture the similarity between these two users we train user embeddings on the set of users  $U \cup\{u_a',u_a''\}\setminus\{u_a\}$.
We generated a test set of 1,000 pairs using this method.
Each training of user embeddings consisted of 1,000 epochs, with shuffling after each epoch.

Results of experiments\footnote{We implemented \textsc{Traj2User} on PyTorch. Upon acceptance we will make the code available on GitHub.} (Table \ref{tab:results}) show that \textsc{Traj2User} outperforms the other tested method by a large margin, i.e., a $14.7\%$ relative improvement over \textsc{sum}, the second best performer\footnote{The difference is statistically significant for a t-test on the reciprocal rank score across the 1,000 test pair with $p=3.66\cdot10^{-5}$.}.
An interesting negative result emerges from the comparison of \textsc{Sum}, with PPMI, SM and their SVD methods.
These latter methods, that are typically applied with success on language modeling tasks perform poorly on our task.
We measured that the distribution of frequency of values in the TagMyDay dataset follow a logarithmic distribution and not the typical Zipf distribution of words in text, yet it is hard to consider this difference as the cause of the drop in MRR. We leave the investigation of this aspect to future work.
Softmax-based methods are by far the worst performers, indicating that forcing a probabilistic interpretation of the observed data is a wrong design choice.
Shorter embeddings, either from truncated SVD or setting smaller size in \textsc{Traj2User}, reduce the MRR.
However, even the shortest \textsc{Traj2User} embedding, with $f=8$ ($|e_i|=11$) outperforms \textsc{Sum}, indicating a graceful degradation of performance and the possibility of exploiting data compression.
Larger embeddings performed as the original-length ones, indicating that the information contained in the relatively small dataset we were able to obtain was already fully modeled in the original-length embeddings.

\begin{figure}
\centering
\includegraphics[width=0.6\columnwidth]{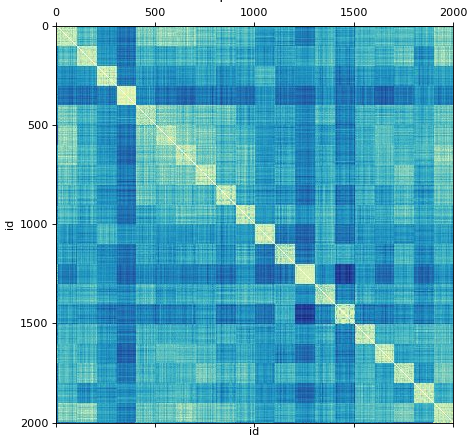}
\caption{\label{fig:virt_sim} User-by-user similarity modeled by \textsc{Traj2User} on a simulated population with 20 groups, each one of 10 users (the lighter the color the most similar the pair of users).}
\vspace{-1em}
\end{figure}

We ran a further experiment to check if \textsc{Traj2User} user embeddings are able to discover groups of similar users and to consistently model similarities across groups.
We used the method for the creation of virtual users described in this section to create a population of 2,000 users composed of 20 groups of 100 virtual users, each one generated from a real user randomly sampled from the TagMyDay dataset. 
Figure \ref{fig:virt_sim} visualizes the cosine similarity among users measured on \textsc{Traj2User} user embeddings. 
Virtual users in the same group are adjacent.
As shown, \textsc{Traj2User} captures the (imposed) within-group similarity between users (the light blocks on the diagonal), and also consistently models the (casual) inter-group similarities.

\section{Conclusions and Future Works}
\label{sec:conclusions}
\textsc{Traj2User} is an innovative way to generate effective user embeddings, starting from simple representations of semantic trajectories and delegating the definition of the similarity model to the learning process of the network.
Although the TagMyDay dataset is limited to a few labels, the method is general enough to expand the amount and forms of semantic information (e.g. interactions of the user with social platforms like ratings and comments on Foursquare or TripAdvisor). 
For example, when the information comes in the form of a piece of text, it can be encoded into a semantic-rich vector using neural language models, e.g., paragraph vectors \cite{LeM14}. As future works we plan to make experiments on other datasets publicly available gathered from social media, and to extend the \textsc{Traj2User} network with a recurrent component, so as to model multi-segment trajectories as a single entity.

\section*{Acknowledgments}
\label{sec:ack}
The present paper has been partially supported by the MASTER project that has received funding from the European Union's Horizon 2020 research and innovation programme under the Marie-Slodowska Curie grant agreement N. 777695 and the FAPESC project MATCH N. UER2017061000002

\bibliographystyle{plain}
\bibliography{biblio}

\end{document}